# Comment on „Isotope effect in multi-band and multi-channel attractive systems and inverse isotope effect in iron-based superconductors" by T. Yanagisawa et al.

Annette Bussmann-Holder and Hugo Keller*
Max-Planck-Institut für Festkörperforschung, Heisenbergstr. 1, D-70569 Stuttgart, Germany
*Physik-Institut der Universität, Winterthurerstr. 190, CH-8057 Zürich, Switzerland

In a recent paper Yanagisawa et al. [1] claim from a theoretical analysis of a multi-channel multi-band superconductor model that an inverse isotope exponent on the superconducting transition temperature $T_c$ can be realized in iron-based superconductors. Simultaneously, a subgroup of the authors of Ref. 1 performed the corresponding isotope effect experiment on $(Ba, K)Fe_2As_2$ by investigating the iron isotope exchange effect on $T_c$ [2]. In accordance with their theoretical analysis they indeed report an unusually large sign reversed isotope exponent of $\alpha \approx -0.18(3)$ which is in strong contrast to previous experiments on the nominally same system with the same composition in Ba, K content, namely $Ba_{0.6}K_{0.4}Fe_2As_2$ [3], where the exponent was determined to be $\alpha \approx 0.37(3)$. This conflict remains unsolved until now with the exception of Ref. 4 where the iron isotope exponent has been determined for FeSe. In accordance with the results of Ref. 3 a large positive isotope exponent has been seen thus questioning the outcome of Ref. 1 and implicitly the findings of Ref. 2. Here, we do not comment on the controversial experimental situation but address the theoretical analysis of Ref. 1, where a variety of misleading assumptions have led to the conclusion that a sign reversed isotope exponent can be realized in a multi-band and multi-channel attractive model for iron based superconductors.

In this comment we derive the exact expressions for the two cases mentioned in the title of Ref. 1 and proof that a reversed isotope exponent is not possible for both scenarios unless unphysical assumptions are being made. Note, that already in 1963 J. Kondo [5] has studied a similar problem as outlined in [1] and arrived at the conclusion that the isotope effect can vanish, but not reverse the sign.

In the single-band multi-channel case studied first, the authors of Ref. 1 assume that the two pairing interactions of different origin operate within a single band and lead to the appearance of two gaps, one being related to a phononic mechanism, the other to an antiferromagnetic (AF) interaction. The effective interactions have specific energy ranges within this single band with a cutoff at $\omega_1$ for the phononic part and a range from $\omega_1$ to $\omega_2$ in the AF channel. This specific choice, which is also used for the two-band case, is the origin of the renormalized $\lambda_2^*$ (equ. 7 in Ref. 1) and $\lambda_{AF}^*$ (equ. 25 in Ref. 1) which – in turn – cause the inverse isotope effect. Obviously, the model is rather unphysical since it assumes that for a limited k-space range phonons cause the electron-electron attraction, directly followed by AF fluctuation mediated pairing starting from the same k-value where the phononic interaction terminates. The resulting gap, which stems from two gaps due to different pairing potentials, is thus continuous in k-space and cannot be viewed as a two-gap model. It is in particular rather amazing that a single electronic band develops pairing correlations stemming from very different potentials. In spite of this very unusual assumption, let us assume that two channels for pairing exist and refrain from authors' assumption on the cut-off energies but use the conventional notation of $V_i(k,k')$ being attractive within a range of $\pm \hbar \omega_i$ (i=1, 2) then the correct equation for $T_c$, using the authors' notation, should read:

$$\ln\left[\frac{C(\omega_1^{\lambda_1}+\omega_2^{\lambda_2})}{k_B T_c}\right] = \frac{1}{\lambda_1+\lambda_2} \qquad (1)$$

where $C$ contains the constants. This equation can be rewritten as:

$$k_B T_c = C\omega_1\left(\frac{\omega_2}{\omega_1}\right)^{\frac{\lambda_2}{\lambda_1+\lambda_2}} \exp[-1/(\lambda_1+\lambda_2)] \qquad (2)$$

Obviously, either a full or a reduced isotope effect on $T_c$ results, or even a vanishing one, in case $\omega_1, \omega_2$ are both stemming from AF fluctuations. However, a sign reversed exponent can be excluded.

The three cases discussed subsequently by the authors are meaningful only in the first case since cases b) and c) imply that no gap appears in the respective *repulsive* channel. Then the $T_c$ equation reduces to the single-gap BCS equation with the prefactor either being given by $\omega_1$ or by $\omega_2$ which leads to the BCS isotope exponent if any of these factors stem from phonons or to a zero isotope effect if AF fluctuations are at work.

Analogous to the single-band multi-channel case discussed above, the authors introduce in the two-band multi-channel model the same artificial cut-off scheme. Accordingly, their $T_c$ equation is substantially modified as compared to a general approach and assumes a rather simple expression (equ. 22). However, without these very specific assumptions a different expression for $T_c$ is obtained. In order to derive this we start from the original Hamiltonian as given in Refs. 5 and 6, namely:

$$H = \sum_k \varepsilon_i(k) c_{k\sigma}^+ c_k - \sum_{k,k'} V_{11}(k,k') c_{1,k\uparrow}^+ c_{1,-k\downarrow}^+ c_{1,-k'\downarrow} c_{1,k'\uparrow}$$
$$- \sum_{k,k'} V_{22}(k,k') c_{2,k\uparrow}^+ c_{2,-k\downarrow}^+ c_{2,-k'\downarrow} c_{2,k'\uparrow} - \sum_{k,k'} V_{12}(k,k') c_{1,k\uparrow}^+ c_{1,-k\downarrow}^+ c_{2,-k'\downarrow} c_{2,k'\uparrow} \qquad (3)$$
$$- \sum_{k,k'} V_{21}(k,k') c_{2,k\uparrow}^+ c_{2,-k\downarrow}^+ c_{1,-k'\downarrow} c_{1,k'\uparrow}$$

where $\varepsilon(k)$ is the momentum $k$ dependent kinetic energy in band $i, j$=1, 2, the *intraband* pairing potentials are given by $V_{ii}$ whereas the *interband* pairing potentials are denoted by $V_{ij}$. The electron creation and annihilation operators ($c^+, c$) are differentiated with respect to the band from which they originate by an additional subscript $i$=1, 2. In analogy to Refs. 5 and 6 and the original BCS concept, the momentum summations extend over energies within a distance of $\pm \hbar \omega_i$ of the Fermi surface. In order to compare with the results from Ref. 1, the same notations as used there are introduced, namely that the pairing potentials stem from different sources in the two channels, one being phonon (ph) mediated whereas the other is due to antiferromagnetic (AF) interactions. Also, effective coupling constants are introduced by replacing the Fermi surface averaged potentials by $\lambda_{ii}^{\alpha\alpha} = \langle V_{ii}^{\alpha\alpha}(k,k')\rangle_{FS}$ where $ii = AF, ph$ and $\alpha\alpha = 1,2$ with similar expressions for the interband terms. Then the coupled gap equations are obtained as:

$$\Delta_i(k) = \lambda_{ii}^{\alpha\alpha}\int_0^{\hbar\omega_i}\frac{\Delta_i(k')}{2E_i(k')}\tanh\left(\frac{E_i(k')}{2k_BT}\right) + \lambda_{ij}^{\alpha\beta}\int_0^{\hbar\omega_j}\frac{\Delta_j(k')}{2E_j(k')}\tanh\left(\frac{E_j(k')}{2k_BT}\right) \quad (4)$$

where $i,j,\alpha,\beta$ have the same meaning as above and $E = \sqrt{\Delta^2 + \varepsilon^2}$. The transition temperature $T_c$ is defined by the condition that both $\Delta_{i,j}$ are simultaneously zero. At this stage we demonstrate the reason of the discrepancy between the model introduced in Ref. 1 and the correct expression based on the original two-band model (Ref. 6). In [1] the integration limits in equ. 4 have been modified such that the AF channel mediated pairing sets in where the ph-channel pairing terminates and is limited at an energy given by $\omega_j = \omega_{AF}$. This particular choice has no acceptable physical origin and yields the artifacts for the isotope effect. By solving equ. 4 for $T_c$ the following result is obtained:

$$k_BT_c = \sqrt{C\omega_{AF}\omega_{ph}}\exp\left[-\frac{1}{2}\frac{\lambda_{AF}^{\alpha\alpha} + \lambda_{ph}^{\alpha\alpha}}{\tilde{\lambda}} \pm \left(\frac{1}{4}\left\{\ln\omega_{ph} + \ln\omega_{AF} + \frac{\lambda_{AF}^{\alpha\alpha} + \lambda_{ph}^{\alpha\alpha}}{\tilde{\lambda}}\right\}^2 - \frac{1}{\tilde{\lambda}}\right)^{1/2}\right] \quad (5)$$

with $\tilde{\lambda} = \lambda_{AF}^{\alpha\alpha}\lambda_{ph}^{\beta\beta} - \lambda_{AF}^{\alpha\beta}\lambda_{ph}^{\beta\alpha}$. This expression is in agreement with the results obtained by Suhl et al. [6], who introduced the two-band model and with those of Ref. 5. (In the square root expression the constants appearing in the logarithm have been set equal 1.) Clearly, a reduced isotope effect on $T_c$ appears as a consequence of the presence of the AF channel. If both channels were of other than phononic origin the isotope exponent is zero (as has also been pointed out in [5]). On the other hand, a BCS exponent results if only phonon mediated interactions are present. What can not happen is a sign reversal as suggested in Ref. 1.

From the above it is obvious that a sign reversed isotope exponent is the consequence of the approximations used in the model of Ref. 1. A more thorough and careful analysis as presented in this comment reveals that the isotope exponent cannot change sign in the scenarios discussed in [1] in spite of the fact that it can become very small or even zero. As such, the reported sign reversed isotope exponent [2] cannot be supported from the above analysis even if spin-fluctuations are the only pairing mediator in iron-based superconductors. This does, however, not imply that the experiment is wrong, but only that other origins than AF fluctuations can be at play to cause the sign reversal. To conclude, the commented paper does not help in clarifying the existing experimental controversy on this issue.